# Thermodynamic Observation of a Vortex Melting Transition in the Fe-based Superconductor Ba$_{0.5}$K$_{0.5}$Fe$_2$As$_2$

H. K. Mak[1#], P. Burger[2,3#], L. Cevey[1], T. Wolf[2], C. Meingast[2] and R. Lortz[1£]

[1]*The Hong Kong University of Science & Technology, Clear Water Bay, Kowloon, Hong Kong*

[2]*IFP, Karlsruhe Institute for Technology, 76021 Karlsruhe, Germany*

[3]*Fakultät für Physik, Karlsruhe Institute of Technology, 76131 Karlsruhe, Germany.*

In cuprate high-temperature superconductors the small coherence lengths and high transition termperatures result in strong thermal fluctuations, which render the superconducting transition in applied magnetic fields into a wide continuous crossover. A state with zero resistance is found only below the vortex melting transition, which occurs well below the onset of superconducting correlations. Here we investigate the vortex phase diagram of the novel Fe-based superconductor in form of a high-quality single crystal of Ba$_{0.5}$K$_{0.5}$Fe$_2$As$_2$, using three different experimental probes (specific heat, thermal expansion and magnetization). We find clear thermodynamic signatures of a vortex melting transition, which shows that the thermal fluctuations in applied magnetic fields also have a considerable impact on the superconducting properties of iron-based superconductors.

## I.   INTRODUCTION

In classical type-II superconductors, typically two distinct superconducting phases are found: the Meissner state in low fields and the Abrikosov state with field-induced vortices or flux lines which form a regular long-range ordered periodic lattice of typically hexagonal arrangement. In the presence of disorder in the crystal structure, flux pinning effects can furthermore prevent the formation of the regular lattice structure and result in a rich variety of various amorphous glassy vortex phases [1-3]. In the layered cuprate high-temperature superconductors with their short coherence lengths and high transition temperatures, critical thermal fluctuations of the 3D-XY universality class have a significant impact on their superconducting properties in a large range around the superconducting transition temperature [4-7]. As a result, the superconducting transition in finite magnetic fields is transformed into a continuous crossover [5-9]. The underlying physics has e.g. been attributed to a magnetic-field-induced finite size effect due to the limitation of the coherence volume by a magnetic length scale closely related to the vortex separation [9]. As a result, the solid vortex phases melt into a liquid vortex phase well below the superconducting transition temperature, which has been observed experimentally in the ReBa$_2$Cu$_3$O$_{7-\delta}$ compounds with Re = Y, Eu, Dy or Nd [10-21]. Signatures of a vortex melting have been furthermore observed in some classical superconductors, but only in close proximity to the superconducting transition [22-25].

---

[#] These authors contributed equally to the article.

[£] Corresponding author: lortz@ust.hk

The vortex melting transition represents the only sharp phase transition in a magnetic field which can be either of a first-order or second order nature , depending on whether the solid phase is crystalline or glassy [10-20,26-32]. It is of particular importance for technical applications of superconductors, as only the solid vortex phase can show true zero resistance. In the liquid state the vortices can move individually and freely, whereas in the solid phase the finite shear modulus causes collective pinning and thus can effectively anchor the vortices and prevent dissipation in presence of a current [3]. In this sense, the vortex melting transition represents the true superconducting transition in applied magnetic fields. Early approaches to describe this transition with a model of a phonon-induced melting [3,33,34] were based on a Lindeman criterion [35], which states that when the amplitude of thermally-induced vortex vibrations about their equilibrium position reaches a critical value the solid lattice becomes unstable and melts. Later, experimental indications have been found that the vortex melting transition is closely linked to the main superconducting transition and may be rather induced by the increasing strength of critical fluctuations upon approaching the superconducting transition [7,23,36-40].

With the recent discovery of a new class of high-temperature iron-based superconductors based on 2D Fe-As layers [41], high-temperature superconductivity research has again become a primary focus of solid-state physics research. These superconductors show many similarities to the cuprate high-$T_c$s but also significant differences [42,43]. Thermal fluctuations have also been observed in the iron-based superconductors and appear to be stronger in the more two-dimensional 1111 compounds [44] than in the 122 compounds [45]. The overall strength of the fluctuations seems to be intermediate between the behavior in the cuprates and more classical superconductors like $Nb_3Sn$ [23,24]. The reasons are certainly the lower transition temperatures and the weaker anisotropy and larger coherence lengths. Careful high-resolution specific-heat measurements nevertheless showed that thermal fluctuations and vortex melting can even be observed in the 3D bulk classical superconductor $Nb_3Sn$ [23,24]. This raises the question whether vortex melting also exists in the Fe-based superconductors. A perfect tool for investigating vortex melting are specific-heat measurements, as the heat capacity - in contrast to other thermodynamic probes - is typically unaffected by irreversible flux pinning effects [24]. These cause irreversible hysteresis effects, e.g. in the magnetization, which may mask the tiny thermodynamic vortex melting anomalies. The specific heat has been proven to be a powerful method to reveal the first-order character of the melting transition in $YBa_2Cu_3O_{7-\delta}$ [10,13,46], $NdBa_2Cu_3O_{7-\delta}$ [21] and in $Nb_3Sn$ [23]. On the other hand, thermal expansion measurements are particularly sensitive to pinning and thus serve as a sensitive probe of the vortex relaxation occurring around a vortex melting transition with some disorder [9]. In this paper we report high-resolution measurements of the specific heat in combination with DC magnetization and thermal expansion experiments on a high-quality single crystal of $Ba_{1-x}K_xFe_2As_2$ with x=0.5 in magnetic fields up to 14 T. All three experimental probes show clear signatures of a sharp transition, which we associate with an underlying vortex melting transition in the presence of weak pinning.

## II.     EXPERIMENTAL

The $Ba_{0.5}K_{0.5}Fe_2As_2$ single crystals were grown from self flux using an $Al_2O_3$ crucible. Ba and K (3:2) were mixed with prereacted FeAs in a ratio of 1:5, filled into the crucible and enclosed and sealed in a steel container. After heating to 1151°C, the crucible was cooled down very slowly to

1051°C at a cooling rate of 0.20°C/h. At the end of the growth the crucible was tilted to decant the remaining flux and then was slowly pulled out of the furnace.

The magnetization, specific heat and thermal expansion measurements have all been performed on pieces taken from the same single crystal. The specific-heat experiments on a tiny crystal with a mass of 60 μg were performed under field-cooled conditions (FC) with a home-made high-resolution micro-calorimeter which is perfectly adapted to measurements of samples with masses down to a few μg. It can be used with either a DC 'long-relaxation' technique for measurements with precision up to 1%, or with a modulated-temperature AC technique. The latter is less accurate, but provides high-resolutions of $\Delta C/C$ of $10^{-5}$ at a high density of data points. The data presented in this paper was measured with the AC technique, but calibrated with data taken by the relaxation technique. The magnetization was measured with a commercial Quantum Design Vibrating-Sample SQUID magnetometer under zero-field-cooled (ZFC) and field-cooled conditions, which can provide extremely high resolutions (better than $10^{-8}$ emu). A capacitive dilatometer was used for the measurements of the thermal expansion. As this technique requires a certain sample length of a few mm in order to obtain sufficient resolution, the experiments were done on a second larger piece of the single crystal with mass of 2.3 mg. For the specific heat and thermal expansion all data has been measured under FC conditions. In all the experiments the sample was oriented with the *c*-axis parallel to the applied field.

## III. RESULTS

Figure 1 shows the total heat capacity in zero field and various magnetic fields up to 14 T. The zero-field superconducting transition is seen as a slightly broadened jump centered at 34.4 K. Application of a magnetic field reduced the transition temperature and additionally broadens the transition. In order to look for the vortex melting transition, the zero-field data has been subtracted as a background in Figure 1(b). The large downturn towards higher temperature is caused by the superconducting transition anomaly at $T_c$ in the subtracted zero field data. Below this downturn, tiny upwards step-like anomalies can be resolved, which move towards lower temperature upon increasing field (arrows in Fig. 1b). The anomalies resemble those observed in the specific heat of $YBa_2Cu_3O_{7-\delta}$ [10,13,46] and $NdBa_2Cu_3O_{7-\delta}$ [21], where the transition appears to be second-order due to residual pinning. For fields higher than 6 T, there appears to be a very small first-order peak superimposed on top of this step, however the effect is close to the resolution limit of our experiment. The lack of a prominent first-order peak, as we will discuss later, is attributed to flux pinning in our sample. No significant difference was found when the data has been taken upon heating or cooling the sample.

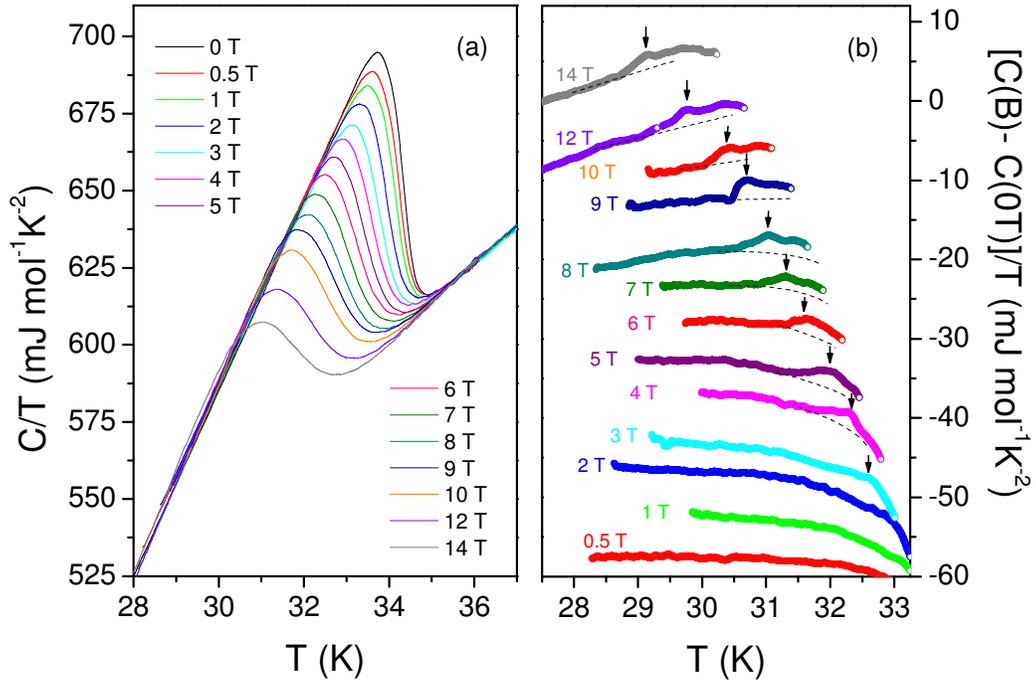

**Figure 1.** (a) Specific-heat of $Ba_{0.5}K_{0.5}Fe_2As_2$ in magnetic fields up to 14 T. (b) The same data as in (a) but with the zero-field data subtracted as a background. The downturn in smaller fields on the right hand side is due to the subtracted zero-field superconducting transition anomaly. Except for the 14 T data, all data has been shifted vertically relatively to each other for clarity. Small steps as marked by the arrows indicate the vortex melting transition anomalies ($T_m$). For clarity, only data in the temperature range of $T_m$ is presented. The dotted lines serve as guides for the eye.

Figure 2 shows data of the magnetization under zero-field-cooled and field-cooled conditions in various magnetic fields. The two branches split at the temperature which is usually defined as the irreversibility temperature $T_{irr}$, below which flux pinning sets in. Above $T_{irr}$, a large reversible range occurs within the superconducting state, which is attributed to the liquid vortex phase. A sharp kink occurs in the reversible part of the curve ($T_{kink}$) slightly above the irreversibility line. As it occurs in the reversible range we attribute it as related to the thermodynamic signature of the vortex melting transition. Both $T_{kink}$ and $T_{irr}$ roughly follow $T_m$ as obtained from the specific-heat data and are therefore likely related to the vortex melting transition.

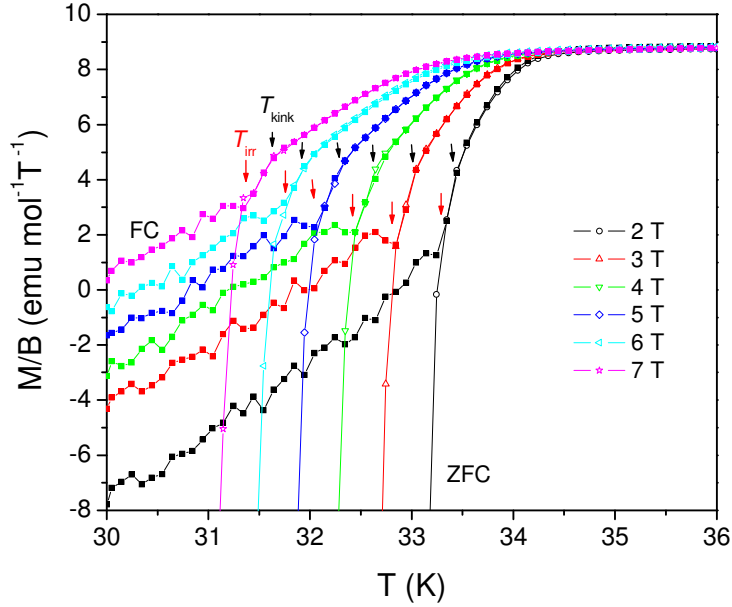

**Figure 2.** DC magnetization data of $Ba_{0.5}K_{0.5}Fe_2As_2$ measured with a VSM SQUID magnetometer upon continuous temperature sweeps upon heating under zero-field cooled and field-cooled conditions. The red arrows mark the irreversibility line ($T_{irr}$) and the black arrows kinks in the reversible magnetization which are attributed to the onset of the vortex melting transition. Note that the positive background originates from the normal-state magnetic properties of the material.

Figure 3 shows linear thermal expansion coefficient $\alpha(T)=1/L_0 \times dL/dT$ measured along the crystallographic $a$-axis upon field cooling (a) and subsequent heating (b). The thermal expansion coefficient $\alpha(T)$ is closely related to the specific heat and phase transitions anomalies appear with the same shape but in contrast to the specific heat can have either a positive or negative signature. The negative signature of the large jump at the superconducting transition indicates, according to the thermodynamic Ehrenfest relation [47], a negative uniaxial pressure dependence of the transition temperature for pressure along the $a$-axis in contrast to the positive pressure dependence reported for Co-doped Ba122 [48,49].

Apart from the large jump at $T_c$, additional smaller anomalies are visible in the heating curves near the proposed vortex melting, as well as a small spike-like anomaly above $T_c$ in both heating and cooling curves. The latter anomaly is, we believe, an artifact, possibly due to a small misaligned crystal region with a slightly higher $T_c$ and it disappears above 3 T. Other crystals from the same batch did not show this feature, and here we concentrate on the anomalies associated with vortex melting, which are observed only upon heating for fields greater than 6 T (see Fig. 6b). In the inset we plot the difference between heating and cooling curves, in which the anomalies at $T_{peak}$ are made more prominent. The absence of these anomalies in the cooling data indicates an irreversible non-thermodynamic origin. The maxima of the peaks coincide rather well with $T_m$ as obtained from the specific heat (see Fig. 1). Similar peaks have been observed at the vortex melting transition of $YBa_2Cu_3O_{7-\delta}$ [50] in samples with weak but finite flux pinning, which were attributed to the decay of some non-equilibrium screening currents near a second-order vortex melting transition [9]. Due to the close similarity of the present effects observed in the thermal expansion with those in YBCO, we also attribute these to an underlying vortex melting transition.

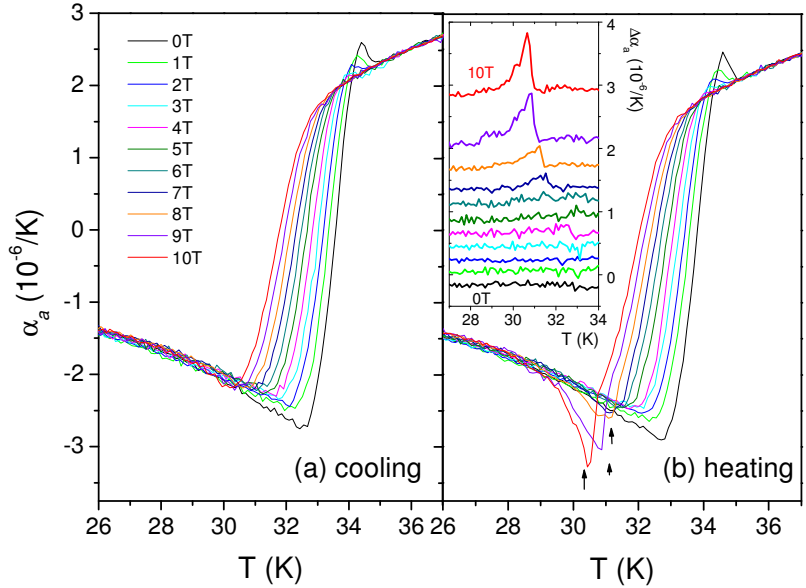

**Figure 3.** Linear thermal expansion data of $Ba_{0.5}K_{0.5}Fe_2As_2$ along the crystallographic *a* axis upon cooling (**a**) and subsequent heating (**b**). Large irreversible peaks ($T_{peak}$) appear upon heating in the vicinity of the vortex melting transition, as marked by the arrows. Inset: Difference $\Delta\alpha$ of the data measured upon cooling and heating to show the peaks more clearly. The date has been shifted vertically relatively to each other for clarity.

## IV. DISCUSSION

Signatures of an underlying vortex melting transition are observed with all three of our experimental probes; the origin of these signatures is however slightly different for each probe as explained in detail below. The specific heat is insensitive to irreversible flux pinning effects [50] and therefore provides a purely thermodynamic signature of the vortex melting transition. The shape of the specific heat transition anomalies shows the characteristic upward step indicative for the additional degrees of freedom in the high-temperature liquid vortex phase. Additionally, in larger fields, a tiny peak, indicative for the latent heat of a first-order melting transition, appears superimposed on this step. In comparison with specific heat on reversible $YBa_2Cu_3O_{7-\delta}$ [10,13,46] and $NdBa_2Cu_3O_{7-\delta}$ [21] samples, the peaks are however much smaller. This suggests that the pinning in our sample is stronger and the crystallization of the flux line lattice is incomplete, presumably related to the formation of amorphous vortex-glass phases [1-3].

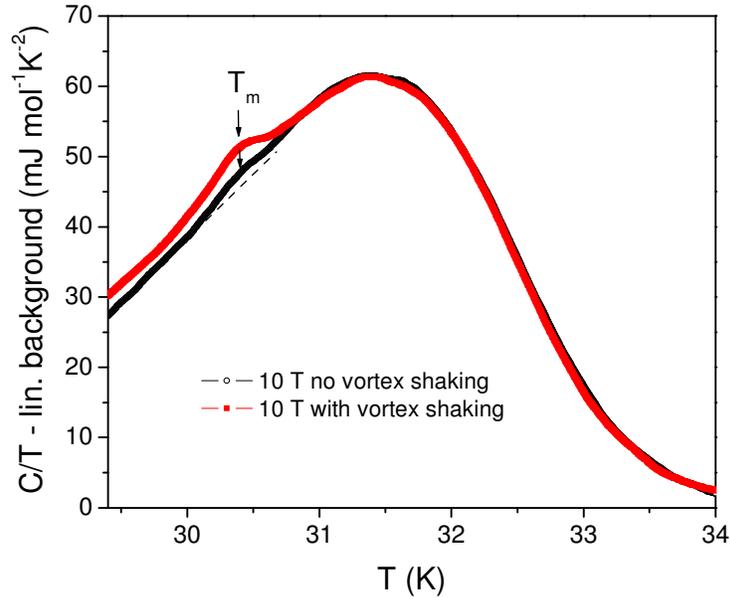

**Figure 4.** Specific-heat data in 10 T with the same linear background fitted above $T_c$ subtracted for clarity. The black data is identical to the data shown in Figure 1. During the measurement of the red data, an additional weak AC magnetic field (~10 Oe, 1 kHz) has been superimposed onto the AC field in order to help the vortices to reach thermodynamic equilibrium ('vortex shaking').

It has been predicted previously that a small additional AC field superimposed onto the applied DC field can help the vortices to overcome the pinning forces and relax into the crystalline state [51,52]. With help of specific-heat data it has been demonstrated e.g. for $Nb_3Sn$ that a vortex melting peak occurs if an AC field of a few Oe in the kHz range is applied parallel to the DC field [23]. In this orientation, the DC field will modulate the density of the vortices and causes vortex vibrations, which help the vortices to overcome pinning forces. To see if the first-order character of the vortex melting transition can be enhanced by such an AC field, we added a small solenoid around our microcalorimeter and applied AC fields of ~10 Oe at frequencies in the kHz range. The result is shown for the 10 T in Figure 4. The AC field causes a strong increase of the specific heat in the vortex melting range and also below. This is indicative for additional degrees of freedom related to unpinned vortices, which can absorb additional heat. Dissipative heating effects from the AC field can be ruled out as the released heat would cause a negative contribution instead of the increase observed here. The peak at the melting transition becomes clearly more pronounced. Although the exact effect of the additional AC field on the superconductor is hard to predict, this experiment shows that the weak pinning does play a significant role in the solid vortex phase and confirms that the solid phase is slightly disordered and most likely represented by the Bragg glass phase predicted by T. Giarmarchi [2].

Magnetization and thermal expansion are probes, which are both sensitive to the reversible thermodynamic contribution, and to irreversible contributions due to flux pinning. The magnetization data in Figure 2 shows a large thermodynamic reversible range above $T_{irr}$, which represents the liquid vortex phase and finally the smooth crossover towards the normal state (which replaces the sharp zero-field superconducting transition in applied fields). The magnetization is however also directly sensitive to induced non-equilibrium screening currents, that appear below $T_{irr}$, as visible in the difference between the ZFC and FC branches. These screening currents may furthermore influence the thermal expansion via the magnetic interaction of the screening currents and the applied field. The applied field applies a 'pressure' on the currents which is transferred to the crystal structure via the flux pinning centers. The observed comparatively large spike-like irreversible anomalies in the thermal expansion in the vicinity of the thermodynamic vortex melting transition are therefore likely related to the decay of such screening currents in the vicinity of the vortex melting transition. The similar anomalies which have been observed previously in $YBa_2Cu_3O_{7-\delta}$ samples in the presence of weak flux pinning [36] have been demonstrated to occur at a second-order vortex melting transition. These anomalies showed behaviour comparable to a kinetic glass transition and are thus most likely related to some glassy vortex phases in the presence of weak collective pinning. Upon approaching the vortex melting transition upon subsequent heating, the currents rapidly decay so that a large peak appears in $\alpha(T)$.

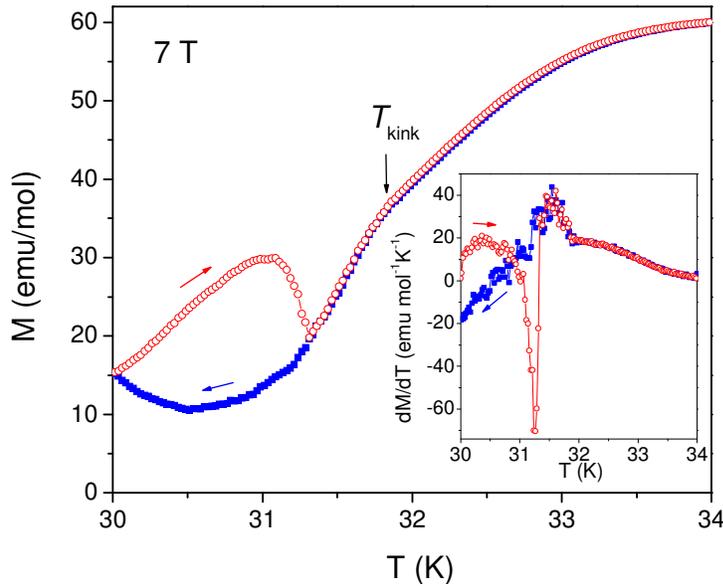

**Figure 5.** FC magnetization data of $Ba_{0.5}K_{0.5}Fe_2As_2$ in a field of 7 T under upon cooling the sample with 0.1 K/min and subsequent heating with 0.1 K/min. A non-equilibrium screening current builds up at low temperature, which decays upon approaching $T_{irr}$. Inset: The corresponding derivative $dM/dT$ of the magnetization. The decay of the induced screening current forms a sharp peak below $T_{irr}$ of identical shape to the peaks occurring in the thermal expansion.

In order to test this scenario and to investigate how these screening currents build up and decay under the field-cooled field-heating conditions of our dilatometry measurements, we performed magnetization measurements under similar conditions (Figure 5). We applied a field of 7 T above the zero-field $T_c$ and then first continuously field-cooled the sample at 0.1 K/min and afterwards heated at the same rate. The two branches are separated below $T_{irr}$ with a small positive contribution appearing in the heating curve, in a similar manner as described by theory in Ref. 53. This difference originates from some non-equilibrium currents (presumably due to temperature gradients in the sample), that are gradually formed when the temperature is changed below $T_{irr}$. In the inset of Figure 5 we show a plot of the corresponding temperature derivative $dM/dT$ of the magnetization. It shows a sharp peak in the heating curve, related to the decay of the screening current when $T_{irr}$ is approached, of very similar shape as in $\alpha(T)$. This confirms that the origin of the sharp peaks in the thermal expansion at the vortex melting transition is due to the decay of screening current due to the abrupt breakdown of collective pinning in the vicinity of the vortex melting transition.

The presence of these peaks further illustrates that our samples are not fully reversible. Thermal expansion data taken on a fully reversible sample of $YBa_2Cu_3O_{7-\delta}$ [36] showed the same reversible thermodynamic signature of the vortex melting transition as the specific heat [46], which is masked in the present heating data by the additional larger irreversible anomalies. The flux pinning, which causing these screening currents, may furthermore explain the small latent heat observed in the specific-heat experiments.

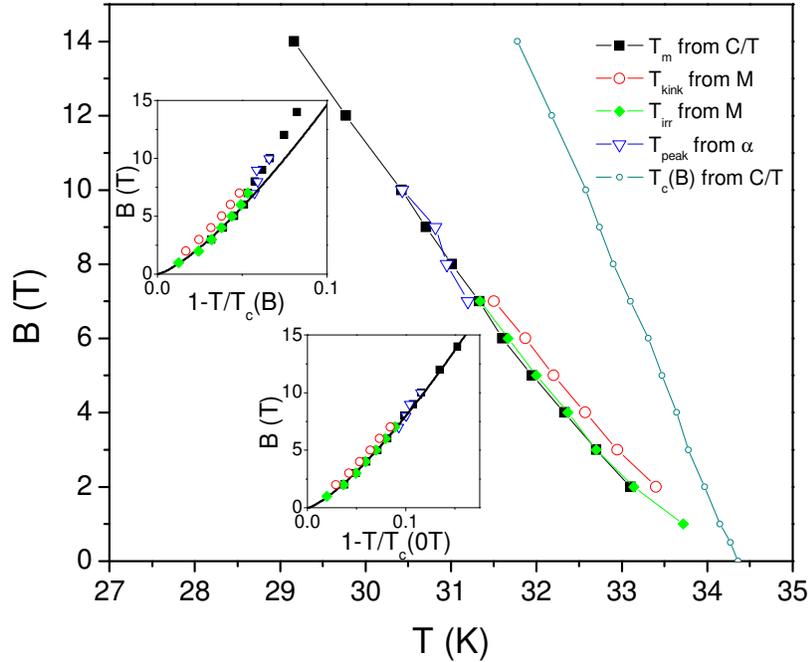

**Figure 6.** Magnetic field versus temperature phase diagram of $Ba_{0.5}K_{0.5}Fe_2As_2$ with the vortex melting transition ($T_m$) as obtained from $C/T$, the irreversibility line ($T_{irr}$), the line where the kink appears in the reversible magnetization ($T_{kink}$), the temperature where the large irreversible peaks appear in the thermal expansivity, as well as the midpoint of the main superconducting transition ($T_c(B)$) in the specific heat. The insets show fits to these data according to the theoretical prediction of a vortex melting transition caused by 3dXY critical fluctuations plotted as function over relative temperature scales $1-T/T_c(B)$ and $1-T/T_c(0T)$ (see text for details).

In Figure 6 we summarize our results in a magnetic field vs. temperature phase diagram of $Ba_{0.5}K_{0.5}Fe_2As_2$. The position of the spike-like anomalies in the thermal expansion coincides well with the position of the thermodynamic signature of the vortex melting transition in the specific heat. Furthermore, the position of the irreversibility line agrees well with the vortex melting transition. The line where the sharp kink occurs in the reversible magnetization runs closely spaced in parallel to the vortex melting transition as extracted from the specific heat. As the step-like anomaly in the specific heat is not perfectly sharp, the kink in the magnetization most likely represents the upper onset of the vortex melting transition. Especially in lower fields all the lines show a typical power-law behavior. In $YBa_2Cu_3O_{7-\delta}$ [13,38] it has been found that the vortex melting line can be described by a power law dependence of the form $B_m \sim (1-T_m/T_c(0T))^{2\nu}$, where $v \approx 0.67$ is a critical exponent of the 3dXY universality class describing the divergence of the coherence length in the vicinity of the critical point at the zero-field $T_c$. This has been dicussed as evidence that the vortex melting is caused by critical fluctuation effects of the 3dXY universality class [7,37-40]. In the insets of Figure 6 we plot the data vs. $1-T/T_c$. The fact that $T_c$ shows a clear field dependence complicates the analysis, as in the pure 3dXY model $T_c$ is not field dependent but only gets broadened by fluctuations. We tried fitting the data both as a function of $1-T/T_c(B)$ and $1-T/T_c(0T)$. When the field-dependent $T_c$ is chosen, the fit fails in the high-field regime. Suprisingly, we can fit the data perfectly by using the zero field $T_c$ instead, which is expected in the 3dXY model. It is however not clear what the implication of this results are, since critical fluctuations in $Ba_{1-x}K_xFe_2As_2$ are certainly expected to be much weaker than in the cuprates. Nevertheless, the fact that the vortex melting line in $Ba_{1-x}K_xFe_2As_2$ follows the same power-law dependence as $YBa_2Cu_3O_{7-\delta}$ indicates that fluctuation effects are of considerable importance for the vortex dynamics in finite magnetic fields of several Teslas.

## V. CONCLUSION

To conclude, we have observed clear signatures of a phase transition in the vortex matter the iron arsenide superconductor $Ba_{0.5}K_{0.5}Fe_2As_2$. The specific heat data are consistent with a second-order type transition, which possibly crossed over to a very weak first-order transition in higher fields. The thermodynamic signature furthermore shows up in the reversible magnetization and is in close vicinity to the irreversibility line below which collective pinning sets in. The linear thermal expansion coefficient shows large spike-like anomalies only upon field-cooled heating, which originate from the rapid decay of non-equilibrium screening currents at the underlying vortex melting transition. They build up continuously upon the preceding field cooling of the sample. The small latent heat at the vortex melting transition in combination with the irreversible effects observed in the magnetization and thermal expansion suggest that the crystallization of the flux line phase is incomplete and the solid phase is mostly likely associated with a weakly disordered topologically ordered Bragg glass phase [2]. The existence of vortex melting in $Ba_{0.5}K_{0.5}Fe_2As_2$ demonstrates that the effect of thermally induced superconducting fluctuations cannot be neglected in iron arsenide superconductors.

**ACKNOWLEDGEMENTS**

We thank Ulrich Welp and Kees van der Beek for useful discussions and U. Lampe for technical support. This work was supported by the Research Grants Council of Hong Kong Grants SEG_HKUST03 and 603010 and by the German priority program on Fe pnictide superconductors (SPP 1458).